\documentstyle[aps,twocolumn,prl,epsf]{revtex}

\newcommand{\be}{\begin{equation}}
\newcommand{\ee}{\end{equation}}
\newcommand{\bea}{\begin{eqnarray}}
\newcommand{\eea}{\end{eqnarray}}

\begin{document}
\draft

 \twocolumn[\hsize\textwidth\columnwidth\hsize\csname @twocolumnfalse\endcsname
\title{Protracted Screening in the Periodic Anderson Model}

\author{A.N.\  Tahvildar-Zadeh$^1$,  M.\ Jarrell $^1$, and J.K.\ Freericks $^2$\\ }
\address{
         $^1$Department of Physics,
University of Cincinnati, Cincinnati, OH 45221\\
         $^2$Department of Physics,
Georgetown University, Washington, DC 20057-0995\\
 	}

\date{\today}
\maketitle
\widetext
\begin{abstract}
\noindent
The asymmetric infinite-dimensional periodic Anderson 
model is examined with a quantum Monte Carlo simulation. For small conduction 
band filling, we find a severe reduction in the Kondo scale, compared to the 
impurity value, as well as protracted spin screening consistent with some 
recent controversial photoemission experiments.  The Kondo screening drives a 
ferromagnetic transition when the conduction band is quarter-filled and both 
the RKKY and superexchange favor antiferromagnetism.  We also find RKKY-driven 
ferromagnetic and antiferromagnetic transitions.
\end{abstract}

]
\narrowtext

\paragraph*{Introduction}  Kondo lattice materials, stoichiometric systems 
generally with U or Ce atoms in the valence shell, have been studied intensely 
over the last few decades.  They display a wide variety of behaviors including 
the well-known transport anomalies associated with a strongly enhanced 
electronic mass, as well as magnetic (antiferromagnetic, incommensurate and 
ferromagnetic), paramagnetic, and superconducting ground states\cite{review}.  
The magnetic nature of the ground state is determined by the strength of the 
hybridization between the f-electrons with the delocalized band
states\cite{Doniach}.  If the hybridization is small, then the RKKY exchange 
dominates and the ground state is magnetic.  If the hybridization is large, 
then either the Kondo screening removes the f-moments, or charge fluctuations 
delocalize the f-electrons destroying their moments, and the ground state 
is a Pauli paramagnet.  

The periodic Anderson model (PAM) is thought to describe the magnetic and 
transport properties of these materials.  Both it and the single impurity 
Anderson model (SIAM) have been extensively studied; nevertheless, the detailed 
nature of the orbitally non-degenerate PAM phase diagram as well as the 
differences in the Kondo screening between the two models are unknown.  In 
this Letter, we present the first rigorous calculations of the phase diagram 
of the asymmetric PAM in infinite dimensions. We find antiferromagnetic, 
ferromagnetic, and paramagnetic ground states, depending upon conduction 
band filling $n_d$ and model parameters.  In the paramagnetic state, we 
find that the temperature dependence of the Kondo screening retains the same 
qualitative features as the SIAM, whereas the {\em{quantitative features 
are quite different}}.  Specifically, when $n_d\ll n_f$ 
($n_d\approx n_f\approx 1$) 
the Kondo scale is strongly suppressed (enhanced) and the temperature 
dependence of the screening is protracted (contracted) compared to the 
SIAM.  These differences between the PAM and SIAM cannot be removed 
by rescaling and may provide insight into recent controversial 
photoemission experiments\cite{jjaa}.

The PAM Hamiltonian is,  
\begin{eqnarray}
H &=& \frac{-t^*}{2\sqrt{D}}\sum_{\langle ij\rangle \sigma}
\left ( d^\dagger_{i\sigma}d_{j\sigma}+{\rm h.c.}\right )\nonumber \\
&+&
\sum_{i\sigma}\left(
\epsilon_{d}d^\dagger_{i\sigma}d_{i\sigma}+
\epsilon_{f}f^\dagger_{i\sigma}f_{i\sigma}
\right)
+V\sum_{i\sigma}\left(d^\dagger_{i\sigma}
f_{i\sigma}+{\rm h.c.}\right)\nonumber \\
&+&\sum_{i} U(n_{fi\uparrow}-1/2)(n_{fi\downarrow}-1/2)\;\;.
\end{eqnarray}
In (1), $d(f)^{(\dagger)}_{i\sigma}$ destroys (creates) a d(f)-electron with
spin $\sigma$ on site $i$ on a hypercubic lattice, $U$ is the screened Coulomb 
repulsion potential for the localized f-states and $V$ characterizes the mixing 
between the two subsystems.  This model retains the complications of the 
impurity problem, which is known to be non-perturbative, including moment 
formation and screening and is further complicated by interactions between
the moments due to RKKY and super-exchange mechanisms.

\paragraph*{Formalism}

A simplifying method which allows for an exact solution of the lattice problem 
in a non-trivial limit is necessary.  Such a method was proposed by Metzner 
and Vollhardt \cite{mevoll} who observed that the irreducible 
self-energy and vertex-functions become purely local as the coordination 
number of the lattice increases. As a consequence, the solution of most 
interacting lattice models may be mapped onto the solution of a local 
correlated impurity coupled to an effective bath that is self-consistently
determined\cite{bramiel90_to_georges92}.  We refer the reader to the above 
references and recent reviews for further details on the method\cite{infdrev}.  

To solve the remaining effective impurity problem, we use the Anderson impurity 
algorithm of Hirsch and Fye \cite{fye}.  In order to model the Ce-based
Kondo lattice materials, we place the correlated f-band below the Fermi
level (so $n_f\approx 1$) and adjust the conduction band filling by varying
the Fermi level.  Thus, beginning at $\beta=10$, we choose $\epsilon_f$ and 
$\epsilon_d$ so that $n_f=1$ and $n_d$ is the desired value.  When the 
temperature is changed, we keep $ \epsilon_f-\epsilon_d$ fixed and vary 
$ \epsilon_f+\epsilon_d$ to conserve the {\em{total}} number of electrons. For 
the results presented here the variation of $n_f$ from one was less than a few 
percent, and statistical error bars are less than 7\%.  Note that our results 
are symmetric under $n_d\to 2-n_d$, and the PAM becomes more correlated for 
small values of $n_d$ and $2-n_d$.

\paragraph*{Results}

Fig.~1 shows the Kondo scale for the PAM and the SIAM versus d-band filling 
when the f-band is half-filled. The Kondo scales are obtained by extrapolation 
$\chi_{imp}(T\rightarrow 0)=1/T_0$, where $\chi_{imp}(T)$ is the additional 
local susceptibility due to the introduction of the effective impurity into 
a host of d-electrons\cite{jarrell}.  We see that at the symmetric limit 
($n_f=n_d=1$) the Kondo scale for the PAM is enhanced compared to the Kondo 
scale of SIAM as has been found earlier \cite{Rice,jarrell}.  However, 
far from the symmetric limit the Kondo scale for the PAM is strongly 
{\em{suppressed}}.

The main consequence of this suppression is that the temperature dependence 
of the screening in the PAM is quite different from that of the SIAM.  This 
is shown in Fig.~2 where the screened local f-moments $T\chi_{ff}$ for both 
models are plotted versus temperature.  As expected from the impurity problem 
$T\chi_{ff}$ displays a log-linear $T$-dependence in some temperature interval.
However, concomitant with the differences in $T_0$, the screening of the 
PAM and the SIAM local moments are quite 
different: for most values of $n_d$ these two curves {\it cannot be made to 
overlap by rescaling their temperature dependencies} (nor is it possible, 
in the PAM case, to make the curves of different d-fillings overlap by 
rescaling $T$).  At high temperatures, the screened local moments are 
identical for the PAM and the SIAM, and as the temperature is lowered below 
$T_{0SIAM}$ the screening begins; however, if $n_d\ll n_f$ ($n_d\approx n_f=1$), 
the screening is significantly protracted (contracted) in temperature for 
the PAM compared to that in the SIAM. (The contracted screening region,
as well as the enhancement of the Kondo scale, when $n_d=n_f=1$, depends 
strongly upon the correlation energy $U$, and diminishes when $U$ is small; 
i.e., for the less correlated data set in Figs.~1 and 2).  This behavior is 
consistent with recent controversial photoemission experiments on 
single-crystals of Ce-based heavy-fermion compounds, where the 
temperature-dependence of the screening peak is much slower than what would 
be expected for a SIAM\cite{jjaa}.  

The Drude weight, $D$, calculated by extrapolation of the current-current 
correlation function\cite{swz}, is shown in the inset to Fig.~1. For all
of the data shown in Fig.~1, $D$ is quite small and the effective electron 
mass (not shown) is large $m^*/m=1/Z \geq 15$, where $Z$ is the quasiparticle 
renormalization factor.  In the symmetric limit, where a gap opens in the 
single-particle density of states, we have $D=Z=0$.  However, consistent with 
what is seen in the Kondo scale for the PAM, $D$ and $Z$ also become very 
small when $n_d\ll 1$.  When the d-filling falls, there are no longer enough 
d-electrons to completely screen the f-moments.  Thus, $T_0$ 
falls quickly as the d-band is doped away from half filling.  If this process 
continues, the d-band becomes depopulated, and then only acts as a hopping 
path between the f-levels.  In the limit, $n_d\to 0$, the PAM may be mapped 
onto a strongly correlated symmetric Hubbard model\cite{TPHMlnca} (with an 
on-site correlation $U$ and a strongly reduced hybridization), which is 
known to open a gap in the single-particle density of states and have $Z=0$ 
and $D=0$\cite{infdrev}.  Thus, when $n_d=0$ or $n_d=1$, we find that both 
the f and the d-band densities of states vanish at the Fermi surface.  
However, for all other values of $n_d$ explored, the f and 
d-density of states remain finite (with the f-DOS only moderately 
enhanced) indicating that the system remains metallic.  

	It is clear from the Drude weight shown in the inset to Fig.~1, 
that the unscreened moments have a dramatic effect on the Fermi-liquid 
properties of the system. This may also be seen by examining the electronic 
distribution function $n(\epsilon_k)=T\sum_n [G^{dd}(\epsilon_k,i\omega_n)+G^{ff}(\epsilon_k,i\omega_n)]$
where $G^{dd}$ and $G^{ff}$ are the fully dressed d- and f-band Green's 
functions calculated with the QMC.  
$d n(\epsilon_k)/d \epsilon_k$, is calculated by
numerically evaluating the derivative of the above sum.  The width of this 
distribution at low temperatures (shown in Fig.~3) gives an estimate of the 
single-particle scattering rate.  It must go to zero if a Fermi-liquid is to 
form.  This appears to happen when $n_d=0.8,0.6$ and $0.4$; however,
for $n_d=0.2$, it is not clear whether a Fermi-liquid forms.  When 
$n_d\alt 0.4$, there is a protracted region in $T$ of strong spin-flip 
scattering, beginning at $T\agt T_{0SIAM}$, and extending down to very low 
temperatures.

In fact, due to magnetic ordering for $n_d\alt0.6$, we find no compelling 
evidence for non Fermi liquid ground states. Fig.~4.\ shows the magnetic 
phase-diagram of the half-filled f-band PAM. $T_c$ is obtained by extrapolating 
or interpolating  the magnetic susceptibility assuming the mean-field form 
$\chi\propto {1\over (T-T_c)}$.  There are two well-known exchange mechanisms 
that are usually responsible for the formation of these magnetic ground-states. 
The superexchange which results from the exchange  of the local f-electrons 
via the hybridization with the d-band; this exchange always favors 
antiferromagnetic order of the half-filled f-band and becomes strongly 
suppressed as the d-band is filled towards $n_d=1$\cite{Levin}.
The  Ruderman-Kittel-Kasuya-Yosida (RKKY) exchange which results from the 
consequent scattering of a d-electron  from two f-moments; 
this exchange varies (in sign and magnitude) as a function of the d-band 
filling\cite{Levin,Fye90}.  The inset to Fig.~4 shows the difference between 
the staggered and 
uniform RKKY exchange, 
$\Delta J_{RKKY}=J_{RKKY}({\bf q}={\bf Q})-J_{RKKY}({\bf q}={\bf 0})$, where 
${\bf Q}$ is the wave-vector corresponding to the corner of the first 
Brillouin zone and
\begin{equation}
J_{RKKY}({\bf q})=-{T\over 2 N} J_{fd}^2  \sum_{n,\bf k} G^{dd}({\bf k},i\omega_n){ G^{dd}({\bf k}+{\bf q},i\omega_n)}\,.
\end{equation}
Here, $J_{fd}=-8V^2/U$, is the effective exchange between the f and
d-band\cite{Schrieffer}. $\Delta J_{RKKY}$ would be proportional to 
the energy gained by the formation of a ferromagnet versus an antiferromagnet 
in the f-band if RKKY were the only exchange mechanism present. 

In the symmetric limit, $n_f=n_d=1$, for the more strongly correlated model
(larger $U$), there is an instability towards antiferromagnetism as expected 
from the sign of the RKKY-exchange.  (The superexchange, which always favors
antiferromagnetism, is expected to be negligible in this limit\cite{Levin}).
For the less correlated model (smaller $U$) the Kondo scale is larger 
(Fig.~1) and the d-electrons effectively screen away the local moments before 
a magnetic transition can occur \cite{Doniach}, so there is no antiferromagnetism 
in these cases.  

As the d-filling decreases from $1$, the transition remains commensurate and
$T_c$ drops quickly to zero.  However, as the system becomes 
more correlated, the size of this antiferromagnetic region increases.   As 
$n_d$ decreases further, the system remains a paramagnetic Fermi liquid 
until the d-band approaches quarter-filling.  Here, both the superexchange 
and the RKKY-exchange still favor antiferromagnetism; nevertheless, the 
system has a ferromagnetic transition.  The rather low transition 
temperature (compared to $T_c$ at the symmetric limit), which is of the 
order of the Kondo scale, indicates that the mechanism behind this 
ferromagnetism might be related to the Kondo screening effect. 

	To see how Kondo screening can lead to a ferromagnetic exchange, 
assume the size of the Kondo polarization cloud is $1/k_F$\cite{barzykin}
so the screening cloud is almost local.  In this situation, every 
time a d-electron hops to the neighboring d-level it breaks its resonance 
with the local f-electron and regains the lost resonance-energy only if 
the f-electron on the neighboring site has the same spin orientation
and the d-level is unoccupied. 

This Kondo-exchange mechanism is substantiated by examining the staggered 
charge susceptibility of the d-levels (not shown) which is strongly 
enhanced at low temperatures near quarter-filling.  Hence, 
each site occupied by a d-electron tends to have its neighboring sites 
unoccupied, optimizing the above mentioned mechanism.  In addition,  
if the quarter-filled d-electrons actually form a spin-polarized staggered 
charge-density wave, where 
the spins of the d-electrons are aligned opposite to the ferromagnetically 
ordered f-electrons, then the quasiparticle band opens a gap at the 
Fermi level. This lowers the kinetic energy of all the occupied d-states.  
The quasiparticle band has to be sufficiently narrow for this 
ferromagnetic charge-density wave to have lower energy than the 
paramagnetically filled quasiparticle band (i.e. the gap must be on the order 
of the quasiparticle bandwidth, $\approx t^*Z\ll t^*$).  Both the charge 
density fluctuation effect and the associated Kondo exchange rapidly 
vanish away from quarter-filling since the charge density wave can 
no longer be commensurate and open a gap at the Fermi level.   Thus, the 
ferromagnetism for smaller values of $n_d$ is due to RKKY exchange (which becomes 
ferromagnetic).  Note that this Kondo-effect generated ferromagnetism at 
quarter-filling is a special feature of the bipartite lattice.

Finally, we speculate that antiferromagnetism should return in the region 
near the empty d-band, since in this limit the half-filled f-band PAM
reduces to a half-filled Hubbard model\cite{TPHMlnca} with 
strong antiferromagnetic superexchange\cite{Levin}.

\paragraph*{Conclusion} 
We investigate the phase diagram  and screening of the 
asymmetric PAM in infinite dimensions. We find antiferromagnetic, 
ferromagnetic, and paramagnetic ground states, depending upon $n_d$ and 
other model parameters.  When the d-band is quarter-filled we find a 
ferromagnetic transition driven by protracted Kondo screening when the 
super and RKKY exchanges are antiferromagnetic.  
In the paramagnetic state, we find that the temperature 
dependence of the Kondo screening retains the same qualitative features as 
the SIAM, whereas the quantitative features are quite different.  Specifically, 
when the number of conduction electrons is significantly less than the number 
of f-electrons, we find that $T_{0PAM}\ll T_{0SIAM}$.  This results in a  
protracted region of spin screening extending from $T\agt T_{0SIAM}$ down 
to $T< T_{0PAM}$ or $T=T_c$.  In this region, the temperature dependence of 
the screened local moment and presumably the associated features such as the 
screening resonance are much slower than those of the SIAM, which may provide 
insight into the temperature dependence of photoemission spectra in some 
Ce-based heavy-fermion lattice compounds\cite{jjaa}.  Finally, it is important 
to note that both the suppression of $T_0$ and the concomitant protracted 
screening disappear when the f-orbital degeneracy $N\to\infty$\cite{sbmft}.  
Thus, $T_{0PAM}$ also depends strongly upon $N$, so the admixture of 
crystal-field split states as well as a larger $N$ could partially restore 
an impurity-like temperature dependence. 

	We would like to acknowledge stimulating conversations with
A.\ Arko,
J.\ Brinkmann,
A.\ Chattopadhyay,
D.L.\ Cox,
M.\ Grioni,
J.\ Joyce, 
Th.\ Pruschke,
Q.\ Si,
and
F.C.\ Zhang.
Jarrell and Tahvildar-Zadeh would like to acknowledge the support of NSF 
grants DMR-9406678 and DMR-9357199. Freericks acknowledges the support of an 
ONR-YIP grant N000149610828.  Computer support was provided by the Pittsburgh 
Supercomputer Center (grant number DMR950010P and sponsored by the NSF)
and the Ohio Supercomputer Center.

\begin{figure}[t]
\epsfxsize=3.5in
\epsffile{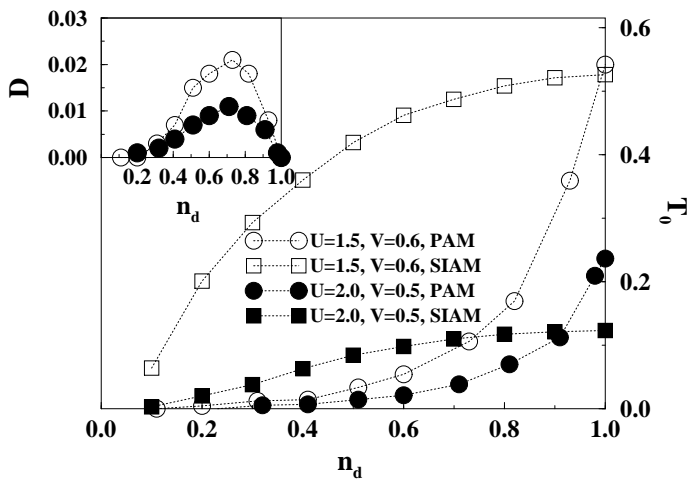}
\caption{ Kondo temperature vs. d-band filling for the infinite-dimensional 
PAM and  SIAM at fixed half-filled f-band ($n_f \approx 1.0$) and two 
different sets of model parameters. The inset shows the corresponding 
Drude weight for the PAM. }
\end{figure}

\begin{figure}[t]
\epsfxsize=3.5in
\epsffile{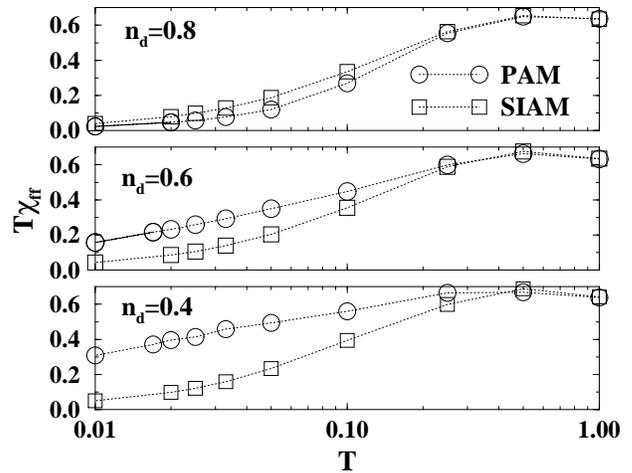}
\caption{ The f-band magnetic moments for the PAM and the SIAM vs. 
temperature for three different values of d-band filling when $U=1.5$ and 
$V=0.6$.}
\end{figure}

\begin{figure}[t]
\epsfxsize=3.5in
\epsffile{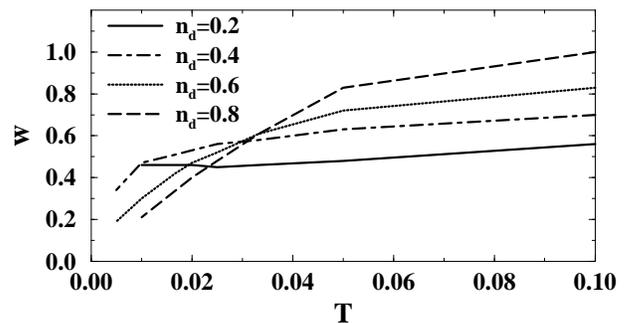}
\caption{ Half-width of the derivative of the total occupation-number, 
${d\ n(\epsilon_k)\over d\ \epsilon_k}$, at low temperatures when $U=1.5$ and 
$V=0.6$. In a Fermi-liquid this quantity should be proportional to $ZT$.  
Any residual value represents the scattering rate at the Fermi surface. The  
flat slope of $w(T)$ at low temperatures for the low-filling case suggests that 
the Fermi-liquid does not begin to form before the ferromagnetic transition 
(cf.\ Fig.~4).}
\end{figure}

\begin{figure}[t]
\epsfxsize=3.5in
\epsffile{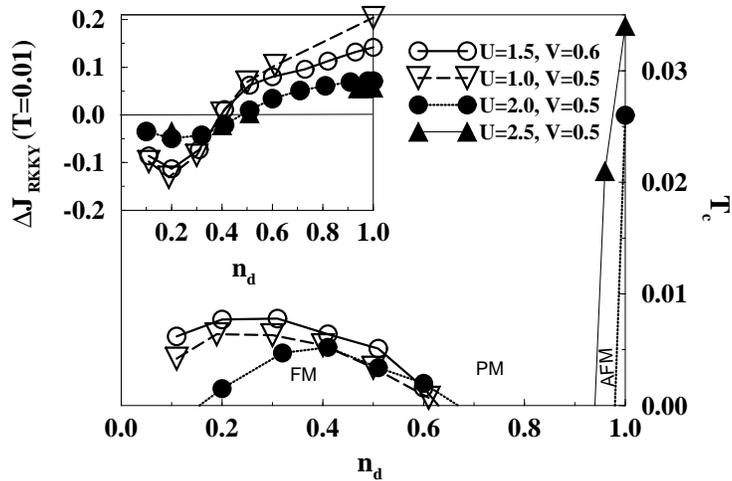}
\caption{ Magnetic phase diagram for infinite-dimensional 
PAM at fixed half-filled f-band 
($n_f \approx 1.0$) and four different sets of model parameters. FM, PM and 
AFM stand for ferromagnetic, paramagnetic and antiferromagnetic phases 
respectively. (The FM transition points for $U=2.5$, $V=0.5$ are not shown.) 
The inset shows the difference between the staggered and uniform  RKKY 
exchange constants vs. d-band filling.  }
\end{figure}

\end{document}